\documentclass[12pt,thmsa]{article}
\usepackage{sw20lart}

\input{tcilatex}
\begin{document}

\title{A CH-type Inequality For Real Experiments}
\author{A. Shafiee\thanks{%
E-mail: shafiee@theory.ipm.ac.ir} \quad \\
{\small \ }$\stackrel{1)}{}${\small Department of Chemistry, Kashan
University,}\\
{\small \ Kashan, 87317-51167, Iran.}\\
$\stackrel{2)}{}${\small Institute for Studies in Theoretical Physics \&
Mathematics,}\\
{\small \ P.O.Box 19395-5531, Tehran, Iran.}}
\maketitle

\begin{abstract}
We derive an efficient CH-type inequality. Quantum mechanics violates our
proposed inequality independent of the detection-efficiency problem.
\end{abstract}

In photonic Bell-type experiments [1], when photon pairs with parallel
linear polarizations are emitted, one can consider a Clauser-Horne (CH)
inequality [2], at the level of hidden variables, in the form

\begin{equation}
-1\leq S_{rq,HV}(\widehat{a},\widehat{b},\widehat{a^{\prime }},\widehat{%
b^{\prime }},\lambda )\leq 0  \tag{1}
\end{equation}
where

\begin{eqnarray}
S_{rq,HV}(\widehat{a},\widehat{b},\widehat{a^{\prime }},\widehat{b^{\prime }}%
,\lambda ) &=&p_{r}^{(1)}(\widehat{a},\lambda )\ \left[ p_{q}^{(2)}(\widehat{%
b},\lambda )-p_{q}^{(2)}(\widehat{b^{\prime }},\lambda )\right]  \nonumber \\
&&+p_{r}^{(1)}(\widehat{a^{\prime }},\lambda )\ \left[ p_{q}^{(2)}(\widehat{b%
},\lambda )+p_{q}^{(2)}(\widehat{b^{\prime }},\lambda )\right]  \nonumber \\
&&-p_{r}^{(1)}(\widehat{a^{\prime }},\lambda )-p_{q}^{(2)}(\widehat{b}%
,\lambda )  \tag{2}
\end{eqnarray}

In (2), we are considering four sub-ensemble of photon pairs with linear
polarizations along $(\widehat{a},\widehat{b})$, $(\widehat{a},\widehat{%
b^{\prime }})$, $(\widehat{a^{\prime }},\widehat{b})$, and $(\widehat{%
a^{\prime }},\widehat{b^{\prime }})$ in which one registers the result $r$
for the first photon with an appropriate probability $p_{r}^{(1)}$ and the
result $q$ for the second one with probability $p_{q}^{(2)}$ where $r,q=\pm
1 $.

To extend the CH inequality to a more efficient one, we propose a function $%
S_{rq}^{\prime }$ in the form:

\begin{eqnarray}
S_{rq,HV}^{\prime }(\widehat{a},\widehat{b},\widehat{a^{\prime }},\widehat{%
b^{\prime }},\lambda ) &=&p_{r}^{(1)}(\widehat{a},\lambda )\ \left[
p_{q}^{(2)}(\widehat{b},\lambda )-p_{q}^{(2)}(\widehat{b^{\prime }},\lambda
)\right]  \nonumber \\
&&+p_{r}^{(1)}(\widehat{a^{\prime }},\lambda )\ \left[ p_{q}^{(2)}(\widehat{b%
},\lambda )+p_{q}^{(2)}(\widehat{b^{\prime }},\lambda )\right]  \nonumber \\
&&-p_{r}^{(1)}(\widehat{a^{\prime }},\lambda )\ p_{r}^{(2)}(\widehat{%
a^{\prime }},\lambda )-p_{q}^{(1)}(\widehat{b},\lambda )\ p_{q}^{(2)}(%
\widehat{b},\lambda )  \tag{3}
\end{eqnarray}
In contrast to the CH inequality, the upper limit of relation (3) should not
necessarily be equal to zero in non ideal experiments. To avoid this
difficulty, we first consider the following inequality for the
single-particle probabilities in an actual experiment:

\begin{equation}
0\leq p_{j}^{(k)}(\widehat{x}_{k},\lambda )\leq 1-p_{0}^{(k)}(\widehat{x}%
_{k},\lambda )  \tag{4}
\end{equation}
where $\widehat{x}_{1}=\widehat{a}$, $\widehat{a^{\prime }}$ or $\widehat{b}$%
, $\widehat{x}_{2}=\widehat{b}$, $\widehat{b^{\prime }}$ or $\widehat{%
a^{\prime }}$ and $j=\pm 1$. The function $p_{0}^{(k)}(\widehat{x}%
_{k},\lambda )$ denotes non-detection probability for the $k$th photon with
the polarization along $\widehat{x}_{k}$. Then, we define the following
relation:

\begin{equation}
\stackunder{j=\pm 1}{\dsum }\ p_{j}^{(k)}(\widehat{x}_{k},\lambda )=\alpha
^{(k)}(\widehat{x}_{k},\lambda )=1-p_{0}^{(k)}(\widehat{x}_{k},\lambda ) 
\tag{5}
\end{equation}
where $\alpha ^{(k)}(\widehat{x}_{k},\lambda )$ is a measure of
inefficiencies at the level of hidden-variables. For more convenience, we
call $\alpha ^{(1)}(\widehat{a},\lambda )\equiv \alpha _{1}$, $\alpha ^{(1)}(%
\widehat{a^{\prime }},\lambda )\equiv \alpha _{1}^{\prime }$, $\alpha ^{(2)}(%
\widehat{a^{\prime }},\lambda )\equiv \alpha _{2}^{\prime }$, $\alpha ^{(2)}(%
\widehat{b},\lambda )\equiv \beta _{2},$ $\alpha ^{(2)}(\widehat{b^{\prime }}%
,\lambda )\equiv \beta _{2}^{\prime }$ and $\alpha ^{(1)}(\widehat{b}%
,\lambda )\equiv \beta _{1}$. According to the definition of the
inefficiency measures in (5), we have:

\begin{eqnarray}
\int_{\Lambda }\alpha ^{(1)}(\widehat{x}_{1},\lambda )\alpha ^{(2)}(\widehat{%
x}_{2},\lambda )\rho (\lambda )d\lambda &=&\ \int_{\Lambda }\stackunder{%
r,q=\pm 1}{\dsum }p_{r}^{(1)}(\widehat{x}_{1},\lambda )p_{q}^{(2)}(\widehat{x%
}_{2},\lambda )\rho (\lambda )d\lambda  \nonumber \\
&=&\stackunder{r,q}{\sum }P_{rq}^{(12)}(\widehat{x}_{1},\widehat{x}%
_{2})\equiv M(\widehat{x}_{1},\widehat{x}_{2})  \tag{6}
\end{eqnarray}
where $P_{rq}^{(12)}(\widehat{x}_{1},\widehat{x}_{2})$ is the joint
probability for getting the results $r$ and $q$ for the first and second
photons along $\widehat{x}_{1}$ and $\widehat{x}_{2}$, respectively, at the
experimental level and $\rho (\lambda )$ is a probability density in space $%
\Lambda $. One can easily show that $M(\widehat{x}_{1},\widehat{x}_{2})=1-$ $%
P_{0}^{(1)}(\widehat{x}_{1})-P_{0}^{(2)}(\widehat{x}_{2})-P_{00}^{(12)}(%
\widehat{x}_{1},\widehat{x}_{2})$ which is a measure of non-detection
probabilities in real experiments. Now, we make the following assumption:

\begin{quote}
\textbf{A}- \textit{The experimental probabilities of non-detection are
independent of the polarization directions.}
\end{quote}

It is important to notice that we are suggesting \textbf{A} only at the 
\textit{observational} level. This indicates that $M(\widehat{x}_{1},%
\widehat{x}_{2})$ should be independent of any direction. Using \textbf{A}
and multiplying the limits of $S_{rq,HV}^{\prime }$ in (3) through $\rho
(\lambda )$ and integrating over the space $\Lambda $, we get the following
inequality at the experimental level:

\begin{equation}
-1\leq S_{rq,\exp }^{\prime }(\widehat{a},\widehat{b},\widehat{a^{\prime }},%
\widehat{b^{\prime }})\leq 0  \tag{7}
\end{equation}

This is our extended CH inequality where $S_{rq,\exp }^{\prime }$ is defined
as follows:

\begin{eqnarray}
S_{rq,\exp }^{\prime }(\widehat{a},\widehat{b},\widehat{a^{\prime }},%
\widehat{b^{\prime }}) &=&P_{rq}^{(12)}(\widehat{a},\widehat{b}%
)-P_{rq}^{(12)}(\widehat{a},\widehat{b^{\prime }})+P_{rq}^{(12)}(\widehat{%
a^{\prime }},\widehat{b})+P_{rq}^{(12)}(\widehat{a^{\prime }},\widehat{%
b^{\prime }})  \nonumber \\
&&-P_{rr}^{(12)}(\widehat{a^{\prime }},\widehat{a^{\prime }})-P_{qq}^{(12)}(%
\widehat{b},\widehat{b})  \tag{8}
\end{eqnarray}

In deriving (7), we have used Bell's locality (factorizability) assumption
[3]. The inequality (7) is violated by quantum mechanical predictions in
non-ideal regime. To show this one can define the joint probability $%
P_{++,QM}^{(12)}(\widehat{a},\widehat{b})$ in a real experiment as [4]

\begin{equation}
P_{++,QM}^{(12)}(\widehat{a},\widehat{b})\approx \frac{1}{4}\eta _{1}\eta
_{2}f\left[ 1+\ F\cos 2(\widehat{a}-\widehat{b})\right]  \tag{9}
\end{equation}
where$\eta _{k}$ ($k=1,$ $2)$ and $f$ are respectively the efficiencies of
the detectors and the collimators, and $F$ is a measure of the correlation
of the two photons. The efficiency parameters in (9) are usually believed to
be independent of the polarization directions in the literature. So, the
assumption \textbf{A} is naturally honored in quantum mechanical
calculations.

Now, we consider the case $\mid \widehat{a}-\widehat{b}\mid =$ $\mid 
\widehat{a^{\prime }}-\widehat{b}\mid =$ $\mid \widehat{a^{\prime }}-%
\widehat{b^{\prime }}\mid =\frac{\varphi }{2}$ and $\mid \widehat{a}-%
\widehat{b^{\prime }}\mid =\frac{3\varphi }{2}$. Substituting (9) and
relations similar to it in (8) and choosing $r=q=+1$, we get

\begin{equation}
S_{++,QM}^{\prime }(\varphi )\approx \frac{1}{4}\eta _{1}\eta _{2}fF\left[
3\cos \varphi -\cos 3\varphi -2\right]  \tag{10}
\end{equation}

Considering the upper limit in (7), we get

\begin{equation}
(3\cos \varphi -\cos 3\varphi )\leq 2  \tag{11}
\end{equation}
This is violated for certain ranges of $\varphi $. We note that none of the
efficiency parameters appear in this inequality. Thus, quantum mechanics
violates (7) independent of the efficiencies of the apparatuses.


\begin{thebibliography}{9}
\bibitem{1}  W. Tittle and G. Weihs, \textit{Quantum Information and
Computation,} \textbf{1}, 3 (2001).

\bibitem{2}  J. F. Clauser and M. A. Horne, \textit{Phys. Rev. D} \textbf{10}%
, 526 (1974).

\bibitem{3}  A. Shimony, in \textit{Sixty-Two Years of Uncertainty:
Historical, Philosophical, and Physical Inquiries into the Foundations of
Quantum Mechanics}, edited by A. Miller\textit{\ }(Plenum, New York, 1990),
pp. 33-43.

\bibitem{4}  . F. Clauser and A. Shimony, \textit{Rep. Prog. Phys}., \textbf{%
41}, 1881 (1978).
\end{thebibliography}
\end{document}